\documentclass[a4paper,10pt]{article}
\usepackage{amsmath}  
\usepackage{amsfonts}
\usepackage{amssymb}
\usepackage{graphicx}
\begin{document}
 \title{Redshift-drift as a test for discriminating between decelerating inhomogeneous and accelerating universe models}
\maketitle 
\author{{\bf Priti Mishra $^{1,a}$, Marie-No\"elle C\'el\'erier $^{2,b}$ and Tejinder P. Singh $^{1,c}$} \\
{\small $^1$ Department of Astronomy and Astrophysics, Tata Institute of Fundamental Research, Homi Bhabha Road, Colaba, Mumbai, 400005, Maharashtra, India} \\
{\small $^2$ Laboratoire Univers et TH\'eories (LUTH), Observatoire de Paris,
CNRS, Universit\'e Paris-Diderot, 5 place Jules Janssen, 92190 Meudon, France} \\
{\small e-mail:  $^a$ priti@tifr.res.in}  \\
{\small $^b$ marie-noelle.celerier@obspm.fr} \\
{\small $^c$  tpsingh@tifr.res.in \\}}
\newcommand\beq{\begin{equation}}
\newcommand\eeq{\end{equation}}
\newcommand\beqa{\begin{eqnarray}}
\newcommand\eeqa{\end{eqnarray}}
\begin{abstract}
Exact inhomogeneous solutions of Einstein's equations have been used in the 
literature to build models reproducing the cosmological data without dark energy. 
However, owing to the degrees of freedom pertaining to these models, it is necessary 
to get rid of the degeneracy often exhibited by the problem of distinguishing between 
them and accelerating universe models. We give an overview of redshift drift in inhomogeneous
cosmologies, and explain how it serves to this purpose. One class of models which fits the 
data is the Szekeres Swiss-cheese class where non-spherically symmetric voids exhibit a 
typical size of about 400 Mpc. We present our calculation of the redshift drift in this model, and compare it with the results obtained by other
authors for alternate scenarios.
\end{abstract}

\section{Introduction}
One test recently proposed in the literature \cite{Uzan08}
to discriminate between different homogeneous and inhomogeneous models is the redhshift-drift.
The redshift-drift is the change in the redshift of a given source observed at 
 different proper times by a comoving observer in an expanding universe [see fig. 1]. 
 \begin{figure}[!htb]
 \begin{center}
 \includegraphics[width=8cm]{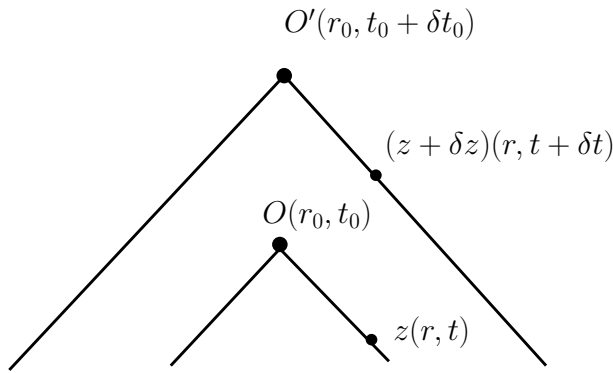}
\caption{The redshift-drift $\delta z$ of a source, initially at a redshift $z$ on the past
 light cone of an observer at $O$, as measured by the same observer at $O'$ after an elapsed 
time $\delta t_0$ of the observer's proper time.}
\end{center}
\end{figure}
Sandage \cite{Sandage62} and then McVittie \cite{MV62} had  first calculated its expression  
in the Friedmannian framework.  In Friedmann-Lema\^itre-Robertson-Walker (FLRW) models, 
 when the Universe expansion decelerates, the redshift 
decreases with time, hence negative redshift-drift and when the expansion accelerates, like in the $\Lambda$CDM model, 
sources with redshifts $\lesssim 2.5$ exhibit a positive redshift-drift. The redshift-drift has been recently calculated for inhomogeneous 
spherically symmetric LTB models  \cite{Yoo11}. We have generalized this calculation to the model proposed in \cite{BC10} and 
derived the equation for the redshift-drift in the axially symmetric QSS model \cite{MCS12}.

 \section{Redshift-drift in quasi-spherical Szekeres models}
 We derive the following equation for the redshift-drift in the axially symmetric QSS model :
\begin{equation}
\frac{{\rm d}}{{\rm d}z} \left(\frac{\delta z}{1 + z}\right) = \frac{1}{(1 + z)^2} \frac{\ddot{\Phi}' - 
\ddot{\Phi}E'/E}{\dot{\Phi}' - \dot{\Phi}E'/E} \delta t_0.
\label{eq19}
\end{equation}
We solve eqn. (\ref{eq19}) numerically to get the redshift-drift ($\dot{z}=\delta z / \delta t_0$).
\begin{figure}[!htb]
\begin{center}
 \includegraphics[width=\textwidth]{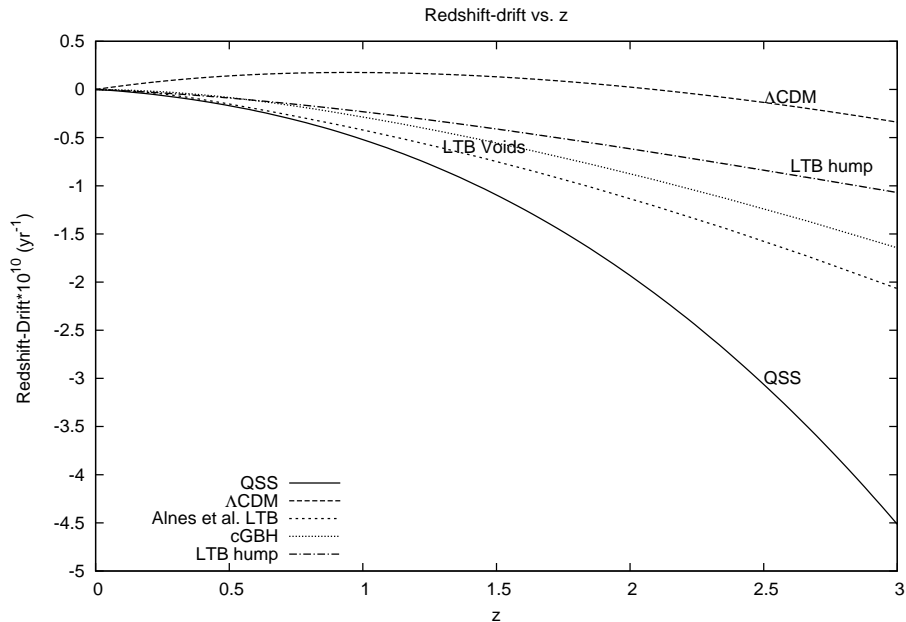}
\caption{The redshift-drift ($\delta z / \delta t_0$) as a function of the redshift $z$ for the axially symmetric QSS 
Swiss-cheese model \cite{BC10}, the $\Lambda$CDM model, the cGBH LTB void model  \cite{garciabellido08} (courtesy \cite{QA10}), 
the Alnes et al. LTB void model \cite{alnes06} (courtesy : \cite{QA10}) and the Yoo LTB hump model \cite{Yoo10} 
(courtesy : \cite{Yoo11})}
\label{drift_compfig}
\end{center}
\end{figure}
In fig. ~\ref{drift_compfig}, we display the redshift-drift for the Szekeres model, the $\Lambda$CDM model, 
and for three LTB models: Alnes et al.'s void model \cite{alnes06} and the constrained GBH (cGBH) void model \cite{garciabellido08}, 
both studied in \cite{QA10}, and Yoo's hump model \cite{Yoo10}, studied in \cite{Yoo11}.

In all three LTB models the observer is located at the center of the void.  The drift for all three LTB as well as for the QSS 
model is negative and its magnitude increases monotonically with the redshift $z$.  The magnitude of the drift in 
the  QSS model is higher by a factor of about two, at a given redshift, than that in the LTB models (at those redshifts where the LTB curves show 
a decline with increasing $z$). 

The drift in the $\Lambda$CDM model is positive up to $z=2.5$, while it is negative in QSS model. At redshift $z=3$, its magnitude 
is much higher than that in the $\Lambda$CDM model by a factor of $\sim 14$,

Thus we have shown that the redshift-drift is a good discriminator between homogeneous and inhomogeneous models.


\begin{thebibliography}{100}


\bibitem{Uzan08}
J.-P. Uzan, C. Clarkson and G. F. R. Ellis, {\it Phys. Rev. Lett}. {\bf 100} 191303 (2008).

\bibitem{Sandage62}
A. Sandage, {\it Astrophys. J.} {\bf 136} 319 (1962).

\bibitem{MV62}
G. McVittie, {\it Astrophys. J.} {\bf 136} 334 (1962).

\bibitem{Yoo11}
C. M. Yoo, T. Kai and K-i Nakao, {\it Phys. Rev}. {\bf D83} 043527 (2011).

\bibitem{BC10}
K. Bolejko and M.-N. C\'el\'erier, {\it Phys. Rev}. {\bf D82} 103510 (2010).

\bibitem{MCS12}
P. Mishra, M.-N. C\'el\'erier and T. P. Singh, {\it Phys. Rev}. {\bf D86} 083520 (2012).


\bibitem{alnes06}
H. Alnes, M. Amarzguioui and \O. Gr\o n, {\it Phys.Rev.} {\bf D73}, 083519 (2006).

\bibitem{garciabellido08}		
J. Garc\'ia-Bellido and T. Haugb\o lle T, {\it J. Cosmol. Astropart. Phys}. JCAP04(2008)003 (2008).

\bibitem{QA10}
M. Quartin and L. Amendola, {\it Phys. Rev}. {\bf D81} 043522 (2010).

\bibitem{Yoo10} C.-M. Yoo, {\it Prog. Theor. Phys}. {\bf 124}, 645 (2010).
\end{thebibliography}
\end{document}